\documentclass[envcountsame,runningheads]{llncs}

\usepackage[utf8]{inputenc}
\usepackage[T1]{fontenc}
\usepackage{lmodern}
\usepackage[hidelinks]{hyperref}
\usepackage{diagbox}
\usepackage{booktabs}
\usepackage{graphicx}
\usepackage{adjustbox}

\usepackage[font=small,labelfont=bf]{caption}
\usepackage{subcaption}
\usepackage{amssymb}
\usepackage{amsmath}
\usepackage{stmaryrd}
\usepackage{tikz}
\usetikzlibrary{arrows.meta, shapes.geometric, fit}
\usetikzlibrary{automata, backgrounds, positioning, arrows, calc}
\usetikzlibrary{decorations.pathreplacing}
\usepackage{mathtools}
\usepackage[linesnumbered,vlined,noend, ruled]{algorithm2e}
\usepackage[inline]{enumitem}  
\usepackage[capitalise]{cleveref}
\usepackage{multirow}
\usepackage{listings}
\usepackage{xcolor}
\usepackage[scaled=.85]{beramono}

 \tikzset{
	module/.style={
		rectangle,
		draw=black!70,
		fill=white,
		thick,
		minimum width=2.4cm,
		minimum height=1cm,
		align=center,
		font=\small\sffamily
	},
	newmodule/.style={
		module,
		draw=black,
		line width=1pt,
		fill=black!8,
	},
	io/.style={
		module,
		rounded corners=3pt,
		fill=white,
		minimum width=2cm,
	},
	flow/.style={
		->,
		>=Stealth,
		thick,
		black!70
	},
	flowlabel/.style={
		font=\scriptsize\sffamily,
		midway,
		fill=white,
		inner sep=1pt
	},
	groupbox/.style={
		rectangle,
		draw=black!40,
		dashed,
		rounded corners=5pt,
		inner sep=8pt
	},
	badge/.style={
		font=\tiny\bfseries\sffamily,
		fill=black,
		text=white,
		rounded corners=2pt,
		inner sep=2pt
	}
}

\newcommand{\oom}{\texttt{m}}
\newcommand{\oot}{\texttt{t}}

\usepackage{todonotes}

\newcommand{\BB}{\ensuremath{\mathbb{B}}\xspace}

\newcommand{\A}{\mathcal{A}}
\newcommand{\B}{\mathcal{B}}
\newcommand{\C}{\mathcal{C}}
\renewcommand{\AA}{\mathbb{A}}
\renewcommand{\BB}{\mathbb{B}}

\usepackage{tikz}
\usetikzlibrary{backgrounds, calc}
\tikzstyle{state}=[thick,minimum size=18pt, circle,draw]
\tikzstyle{transition}=[->,thick,>=stealth,shorten >=1pt,shorten <=1pt, font=\small]
\tikzstyle{loop above right}=[out=60,in=30, min distance=5mm, looseness=8]
\tikzstyle{loop above left}=[out=150,in=120, min distance=5mm, looseness=8]
\tikzstyle{loop below left}=[out=-120,in=-150, min distance=5mm, looseness=8]
\tikzstyle{loop below right}=[out=-30,in=-60, min distance=5mm, looseness=8]
\newcommand{\zokiInitialLength}{18pt}
\newcommand{\zokiInitialAngle}{180}
\newcommand{\zokiInitialPos}{left}
\newcommand{\zokiInitialText}{}
\tikzoption{initial length}{\tikzaddafternodepathoption{\renewcommand{\zokiInitialLength}{#1}}}
\tikzoption{initial angle}{\tikzaddafternodepathoption{\renewcommand{\zokiInitialAngle}{#1}}}
\tikzoption{initial text}{\tikzaddafternodepathoption{\renewcommand{\zokiInitialText}{#1}}}
\tikzoption{initial pos}{\tikzaddafternodepathoption{\renewcommand{\zokiInitialPos}{#1}}}
\tikzstyle{initial}=[after node path={{%
		[to path={[transition, shorten >=3pt] (\tikztostart) --  (\tikztotarget) \tikztonodes}]%
		(\tikzlastnode.\zokiInitialAngle)++(\zokiInitialAngle:\zokiInitialLength) edge[at start] node[\zokiInitialPos, inner sep=0pt, outer sep=1pt] (zokiInitialTextNode) {\zokiInitialText} (\tikzlastnode)%
}}]
\tikzstyle{final}=[after node path={ node[state, scale=.8] at (\tikzlastnode) {} }]
\tikzset{
	bg/.default={},
	bg/.style={execute at end picture={
			\begin{scope}[on background layer]
				\node[xshift=-1mm, yshift=-1mm] (sw) at (current bounding box.south west) {};
				\node[xshift=1mm, yshift=1mm] (ne) at (current bounding box.north east) {};
				\node[xshift=1mm, yshift=-1mm] (nw) at (current bounding box.north west) {};
				\fill[fill=black!10,rounded corners] (sw) rectangle (ne);
				
				\ifx&#1&\else
				\node[anchor=north east, xshift=2pt] at (nw) {#1};
				\fi
			\end{scope}
	}},
}

\newcommand{\ValFunction}[1]{{\textsf{#1}}\xspace}

\newcommand{\Min}{\ValFunction{Min}}
\newcommand{\Max}{\ValFunction{Max}}

\newcommand{\Inf}{\ValFunction{Inf}}
\newcommand{\Sup}{\ValFunction{Sup}}
\newcommand{\LimInf}{\ValFunction{LimInf}}
\newcommand{\LimSup}{\ValFunction{LimSup}}
\newcommand{\LimInfAvg}{\ValFunction{LimInfAvg}}
\newcommand{\LimSupAvg}{\ValFunction{LimSupAvg}}
\newcommand{\SumPlus}{\ValFunction{Sum$^+$}}
\newcommand{\SumMinus}{\ValFunction{Sum$^-$}}
\newcommand{\SumBound}{\ValFunction{Sum$^B$}}

\newcommand{\AlphaExample}[1]{\texttt{#1}}
\newcommand{\req}{\AlphaExample{r}}

\newcommand{\gra}{\AlphaExample{g}}

\newcommand{\other}{\AlphaExample{o}}

\newcommand{\CompClass}[1]{{\textsc{#1}}\xspace}
\newcommand{\PTime}{\CompClass{PTime}}

\newcommand{\PSpace}{\CompClass{PSpace}}

\newcommand{\ExpSpace}{\CompClass{ExpSpace}}

\newcommand{\suchthat}{\;\ifnum\currentgrouptype=16 \middle\fi|\;}

\title{Extending QuAK with Nested Quantitative Automata}
\titlerunning{Extending QuAK with Nested Quantitative Automata}

\author{
	Thomas~A.~Henzinger\inst{1}\orcidID{0000-0002-2985-7724}
	\and 
	Nicolas~Mazzocchi\inst{2}\orcidID{0000-0001-6425-5369}
	\and 
	N.~Ege~Sara\c{c}\inst{3}\thanks{Corresponding author.}\orcidID{0009-0000-2866-8078}
	\and
	Harun~Yılmaz\inst{4}\orcidID{0009-0006-8132-2002}
}
\authorrunning{T.~A.~Henzinger \and N.~Mazzocchi \and N.~E.~Sara\c{c} \and  H.~Yılmaz}

\institute{
	Institute of Science and Technology Austria (ISTA), Austria\\
	\email{tah@ista.ac.at}
	\and
	Slovak University of Technology in Bratislava, Slovak Republic\\
	\email{nicolas.mazzocchi@stuba.sk}
	\and
	CISPA Helmholtz Center for Information Security\\
	\email{ege.sarac@cispa.de}
	\and
	Sabancı University\\
	\email{harun.yilmaz@sabanciuniv.edu}
}
\date{}

\begin{document}
\maketitle

\begin{abstract}
	Quantitative automata (QAs) extend finite-state automata on infinite words with weighted transitions to specify quantitative system properties.
	However, their finite weight sets rule out properties like average response time, where response times can be arbitrarily large.
	Nested quantitative automata (NQAs) overcome this limitation: a parent automaton spawns child automata to compute unbounded values over finite infixes and aggregates them into a final result.
	Despite this expressiveness, NQAs have lacked practical tool support to date.
	
	We close this gap by extending the Quantitative Automata Kit (QuAK), a software tool for QA analysis, to support NQAs.
	Our core contribution is implementing a suite of flattening procedures that reduce NQAs to QAs, leveraging QuAK's existing decision procedures.
	These reductions preserve the answers to threshold decision problems, while allowing users to specify properties in the more expressive NQA formalism.
	The tool handles all combinations of parent aggregators (including limits and averages) and child functions (extrema and monotonic or bounded summations) for which emptiness and universality are known to be decidable.
	Experiments on response-time and resource-consumption benchmarks demonstrate QuAK's effectiveness.
\end{abstract}

\section{Introduction}
\label{sec:introduction}

Formal verification has traditionally focused on boolean properties: a system either satisfies a specification or it does not.
Yet many critical requirements are inherently quantitative; we may require that a server's average workload stays above 20\%, or that a process's peak memory usage stays under 1GB.
Quantitative automata (QAs)~\cite{DBLP:journals/tocl/ChatterjeeDH10} address this need by assigning rational weights to transitions in finite-state $\omega$-automata and aggregating them via value functions such as $\Sup$, $\LimSup$, or $\LimSupAvg$.
The Quantitative Automata Kit (QuAK)~\cite{DBLP:conf/isola/ChalupaHMS24,DBLP:conf/tacas/ChalupaHMS25} provides the first comprehensive tool support for analyzing such automata, implementing algorithms for emptiness, inclusion, and safety-liveness analyses.

However, QAs cannot express a fundamental class of quantitative properties.
Consider \emph{average response time}: a system receives requests and issues grants, and we wish to measure the long-run average delay between each request and the following grant.
For the QA value functions considered in this paper, a fixed finite transition-weight set bounds all values a QA can produce.
Average response time, however, requires unbounded per-event values, since a grant may arrive arbitrarily far after its request.
Therefore, no QA can express such measurements~\cite{DBLP:journals/tocl/ChatterjeeHO17}.

\begin{figure}[t]
	\centering
	\includegraphics[width=\textwidth,alt={Nested quantitative automaton for average response time. The parent automaton spawns a child automaton on each request. The child counts one unit per input position until a grant is read and then returns the accumulated response time. An example prefix shows child return values 5 and 2, while a third child is still active.}]{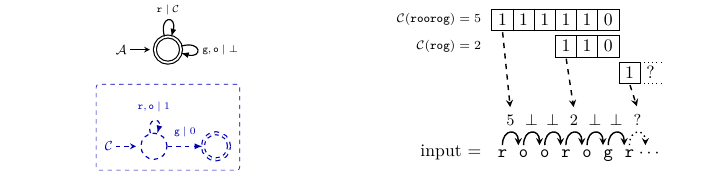}
	\caption{
		A $(\LimSupAvg, \SumPlus)$-automaton $\AA = (\A, \C)$ for average response time.
		\emph{Left}: Parent $\A$ spawns child $\C$ on each request. Child $\C$ accumulates weight $1$ per step until a grant, then terminates. 
		\emph{Right}: On the shown prefix, the spawned instances of $\C$ return $5$ and $2$; the third instance is still active.
		Silent transitions (marked $\bot$) do not contribute to the parent's value, so the running average after the event $\gra$ is $\frac{7}{2}$.}
	\label{fig:response_time_intro}
\end{figure}

Nested quantitative automata (NQAs)~\cite{DBLP:journals/tocl/ChatterjeeHO17} solve this problem.
A \emph{parent} automaton reads an infinite word and spawns \emph{child} automata on designated transitions; each child runs over a finite infix and returns a value upon termination.
The parent aggregates these values using its value function.
\Cref{fig:response_time_intro} shows a $(\LimSupAvg, \SumPlus)$-automaton $\AA = (\A, \C)$ that computes average response time.
The parent $\A$ spawns the child $\C$ on every request; other letters do not contribute a value.
The child $\C$ accumulates weight $1$ on each step until the first grant, then terminates with the accumulated sum.
On the input prefix $\req \other \other \req \other \gra \req \cdots$, the first child returns $5$ (waiting five steps for its grant) and the second returns $2$, while the third is still running.
The parent's $\LimSupAvg$ aggregates these returned values into their long-run average.
Because children can run arbitrarily long, their return values are unbounded, enabling measurement of quantities that non-nested automata cannot express.

Despite solid theoretical foundations and decidability results~\cite{DBLP:journals/tocl/ChatterjeeHO17,DBLP:conf/mfcs/ChatterjeeHO16,DBLP:conf/sas/ChatterjeeHO16}, there has been no practical implementation of NQA algorithms.
We bridge this gap by extending QuAK to support NQAs. 	
In particular:
\begin{enumerate}
	\item
	We extend QuAK with a modular architecture for NQA analysis built around \emph{flattening}: for a given threshold decision query, QuAK translates the NQA instance into a QA instance whose yes/no answer coincides with the original query.
	\item
	We implement flattening procedures for all emptiness and universality problems known to be decidable for combinations of parent aggregators
	$\{\Sup$, $\Inf$, $\LimSup$, $\LimInf$, $\LimSupAvg$, $\LimInfAvg\}$ and child aggregators $\{\Min$, $\Max$, $\SumBound$, $\SumPlus$, $\SumMinus\}$.
	\item
	We evaluate QuAK on two benchmark families---response time and resource consumption---to show its scalability across a range of NQA types and sizes.
\end{enumerate}

Note that these flattening operations need not preserve semantic equivalence between the input NQA and the resulting QA, since NQAs are strictly more expressive than QAs in general, but they preserve the answer to the given threshold query.
Moreover, the benchmarks reflect specifications that arise in protocol latency analysis, service-level response-time monitoring, and resource-consumption analysis for systems with dynamically started and terminated tasks.
In these settings, each request, job, or process lifetime has a finite cost, while the parent captures the worst-case or long-run aggregate of these costs over an infinite execution.

\subsubsection{Overview}
The rest of the paper makes the flattening-based view explicit. Section~\ref{sec:theory} recalls NQAs and their decision problems, Section~\ref{sec:tool} describes QuAK's implementation of the flattening pipeline, and Section~\ref{sec:experiments} evaluates the tool.

\subsubsection{Related Work}
NQAs were introduced in~\cite{DBLP:journals/tocl/ChatterjeeHO17}, which established their expressiveness beyond QAs and much of the decidability landscape.
Bounded-width NQAs~\cite{DBLP:conf/mfcs/ChatterjeeHO16} restrict the number of concurrently active children, yielding better complexity in several cases.
Quantitative monitor automata~\cite{DBLP:conf/sas/ChatterjeeHO16} provide a counter-based formulation expressively equivalent to bounded-width NQAs.
QuAK~\cite{DBLP:conf/isola/ChalupaHMS24,DBLP:conf/tacas/ChalupaHMS25} is the first tool to automate the analysis of QAs; we build directly on its codebase.
Weighted automata tools such as Vaucanson~\cite{DBLP:conf/wia/LombardyPRS03}, Vcsn~\cite{DBLP:conf/wia/DemailleDLS13}, and Awali~\cite{Awali2.3} primarily target finite words with semiring-style semantics.
Signal temporal logic tools (e.g., Breach~\cite{DBLP:conf/cav/Donze10}, S-TaLiRo~\cite{DBLP:conf/tacas/AnnpureddyLFS11}, and RTAMT~\cite{DBLP:conf/atva/Nickovic020}), timed-automata and hybrid-systems tools (e.g., UPPAAL~\cite{DBLP:journals/sttt/LarsenPY97} and HyTech~\cite{DBLP:conf/hybrid/HenzingerH94a}), and probabilistic model checkers (e.g., PRISM~\cite{DBLP:conf/cpe/KwiatkowskaNP02} and Storm~\cite{DBLP:journals/sttt/HenselJKQV22}) address orthogonal quantitative verification problems.

\section{Theoretical Background}
\label{sec:theory}

\subsubsection{Quantitative Automata and Value Functions}
A \emph{quantitative automaton} (QA) over (finite or infinite) words is a tuple $\mathcal{A} = (\Sigma, Q, q_0, \delta, F, C)$ consisting of a finite alphabet $\Sigma$, a finite state set $Q$, an initial state $q_0$, a transition relation $\delta \subseteq Q \times \Sigma \times Q$, an accepting-state set $F \subseteq Q$, and a weight function $C : \delta \to \mathbb{Q}$.

A \emph{run} of $\mathcal{A}$ on a (finite or infinite) word $w = w_1 w_2 \ldots$ is a sequence of states $\pi = q_0 q_1\ldots$ with $(q_{i-1}, w_i, q_i) \in \delta$ for all $i \geq 1$.
Each run induces a weight sequence $C(\pi) = C(q_0, w_1, q_1), C(q_1, w_2, q_2), \ldots$
A \emph{value function} aggregates the weight sequence of a run into a single value.
For infinite sequences, we consider:
\[
\begin{array}{r@{\;=\;}l@{\qquad}r@{\;=\;}l}
	\Inf(x) & \inf_{i \geq 1} x_i & \Sup(x) & \sup_{i \geq 1} x_i \\[2pt]
	\LimInf(x) & \liminf_{i \to \infty} x_i & \LimSup(x) & \limsup_{i \to \infty} x_i \\[2pt]
	\LimInfAvg(x) & \liminf_{k \to \infty} \frac{1}{k} \sum_{i=1}^{k} x_i &\!\!\!\!\!\!\! \LimSupAvg(x) & \limsup_{k \to \infty} \frac{1}{k} \sum_{i=1}^{k} x_i
\end{array}
\]
For finite sequences $x = x_1 \ldots x_n$, we consider $\Min(x) = \min_i x_i$, $\Max(x) = \max_i x_i$, $\SumPlus(x) = \sum_{i=1}^n |x_i|$, $\SumMinus(x) = -\sum_{i=1}^n |x_i|$, and $\SumBound(x)$, which equals $\sum_{i=1}^n x_i$ if all partial sums remain in $[-B, B]$ for a fixed bound $B$, and the first bound crossed ($-B$ or $B$) otherwise.
For a QA $\mathcal{A}$ on infinite (resp. finite) words with value function $f$, a run $\pi$ is \emph{accepting} if it visits $F$ infinitely often (resp. if it ends at a state in $F$), and the \emph{value} of $\mathcal{A}$ on a word $w$ is the supremum of $f(C(\pi))$ over all accepting runs $\pi$ on $w$.
An automaton is \emph{deterministic} if each state has at most one outgoing transition per letter.

\subsubsection{Nested Quantitative Automata}
A \emph{nested quantitative automaton} (NQA)~\cite{DBLP:journals/tocl/ChatterjeeHO17} $\AA = (\mathcal{A}, \mathcal{B}_1, \ldots, \mathcal{B}_k)$ consists of
a \emph{parent automaton} $\mathcal{A} = (\Sigma, Q, q_0, \delta, F, L)$, where $(\Sigma, Q, q_0, \delta, F)$ is an automaton over infinite words and $L : \delta \to \{\bot, 1, \ldots, k\}$ is a labeling function, 
and $k$ \emph{child automata} $\mathcal{B}_1, \ldots, \mathcal{B}_k$, each a QA over finite words.
A parent transition $t$ with $L(t) = j \in \{1, \ldots, k\}$ \emph{invokes} (or \emph{spawns}) child $\mathcal{B}_j$; a transition with $L(t) = \bot$ is \emph{silent} and invokes no child.
An NQA is an \emph{$(f, g)$-automaton} if the parent uses the value function $f$ and every child uses $g$.

A \emph{run} of $\AA$ on $w \in \Sigma^\omega$ is a tuple $(\Pi, \pi_1, \pi_2, \ldots)$ where $\Pi = \Pi[0]\Pi[1]\ldots$ is a run of the parent on $w = w_1 w_2 \ldots$, and for each $i \geq 1$: if the $i$-th parent transition has label $j \in \{1, \ldots, k\}$, then $\pi_i$ is a finite run of $\mathcal{B}_j$ on the finite infix $w_i w_{i+1} \cdots w_{i'}$ of the same input word, for some endpoint $i' \geq i$; otherwise $\pi_i$ is undefined.
Here the $i$-th parent transition reads $w_i$, so the child spawned by that transition starts on the same input letter.
When $\pi_i$ is defined, the child invoked at position $i$ terminates at position $i'$, returning the value $g(C(\pi_i))$.
The endpoint $i'$ is part of the nondeterministic choice of the NQA run: a child spawned at position $i$ may terminate at any position $i' \geq i$ at which it has an accepting finite run on $w_i \cdots w_{i'}$.
The run $(\Pi, \pi_1, \pi_2, \ldots)$ is \emph{accepting} if:
(i) the parent run $\Pi$ visits $F$ infinitely often,
(ii) every invoked child run is finite and accepting, and
(iii) infinitely many parent transitions are non-silent.
In particular, every spawned child must be assigned a finite run ending in an accepting state; a parent run that leaves some child running forever is not accepting.
The \emph{value} of an accepting run is $f$ applied to the sequence of returned child values, omitting silent transitions.
The value of $\AA$ on $w$ is the supremum over all accepting runs on $w$.
An NQA is \emph{deterministic} if the parent and all children are deterministic, and accepting states in each child have no outgoing transitions.
Thus, once a deterministic child reaches an accepting state, its termination position is forced; it terminates at the first accepting position reached by its unique run.
\Cref{fig:response_time_intro} shows a deterministic $(\LimSupAvg, \SumPlus)$-automaton computing average response time.

\subsubsection{Decision Problems}
Given a QA or NQA and a threshold $\lambda \in \mathbb{Q}$, the \emph{emptiness} problem asks whether there exists $w \in \Sigma^\omega$ with value at least $\lambda$, and the \emph{universality} whether every $w \in \Sigma^\omega$ has value at least $\lambda$.
\Cref{tab:complexity} summarizes the complexity landscape from~\cite{DBLP:journals/tocl/ChatterjeeHO17}.
QuAK supports all combinations that are known to be decidable.

\begin{table}[t]
	\centering
	\small
	\renewcommand{\arraystretch}{1.15}
	\caption{Complexity of decision problems for NQAs with parent aggregator $f$ and child aggregator $g$~\cite{DBLP:journals/tocl/ChatterjeeHO17}. $^\dagger$The $(\LimInfAvg, \SumPlus)$ case is open; the others are in \ExpSpace.} \label{tab:complexity}
	\begin{tabular}{@{}l@{\quad}cc@{\quad}cc@{}}
		\toprule
		& \multicolumn{2}{c}{$f \in \{\Inf, \Sup, \LimInf, \LimSup\}$} & \multicolumn{2}{c}{$f \in \{\LimInfAvg, \LimSupAvg\}$} \\
		\cmidrule(lr){2-3} \cmidrule(lr){4-5}
		 & Emptiness & Universality & Emptiness & Universality \\
		\midrule
		$g \in \{\Min$, $\Max$, $\SumBound\}$ & \textsc{PSpace} & \textsc{ExpSpace} & \textsc{PSpace} & Undecidable \\
		$g \in \{\SumPlus$, $\SumMinus\}$ & \textsc{PSpace} & \textsc{ExpSpace} & \textsc{ExpSp}./Open$^\dagger$ & Undecidable \\
		\bottomrule
	\end{tabular}
\end{table}

\subsubsection{Algorithmic Overview}
The decision procedures reduce NQA threshold problems to QA threshold problems via \emph{flattening}: constructing a QA that simulates the relevant information about active child states.
We give the key ideas below; see~\cite{DBLP:journals/tocl/ChatterjeeHO17} for the details.

\textit{(1) Regular children.}
For children with $g \in \{\Min, \Max, \SumBound\}$, the set of possible return values is finite, so for each value $v$, the words of value $v$ form a regular language.
This enables a reduction from $(f, g)$-automata to \emph{semantically equivalent} $f$-automata with silent transitions~\cite[Lem.~4.10]{DBLP:journals/tocl/ChatterjeeHO17}: for each child invocation guess the child's return value, verify the guess via a DFA recognizing the corresponding language, and emit the guessed value as the transition weight.
The construction incurs exponential blowup in the total size of these DFAs.

\textit{(2) Threshold-based bounding.}
For NQAs with $f \in \{\Sup, \LimSup, \Inf, \LimInf\}$ and $g \in \{\SumPlus, \SumMinus\}$, child return values exceeding a given threshold $\lambda$ can be truncated without affecting the emptiness or universality decision~\cite[Thm.~4.18]{DBLP:journals/tocl/ChatterjeeHO17}, reducing to the $\SumBound$ case.

\textit{(3) Extremal parents with monotonic children.}
For emptiness checking of NQAs with $f \in \{\Sup, \LimSup, \Inf, \LimInf\}$ and $g \in \{\Min, \Max, \SumPlus, \SumMinus\}$, we provide specialized guess-and-verify procedures that exploit two properties to achieve smaller state spaces than the general reduction.
First, the child functions are \emph{monotonic}: for $\Min$ and $\Max$, we track the closest value to a guessed return value seen so far; for $\SumPlus$ and $\SumMinus$, the guessed value acts as a budget to produce or consume.
This enables pruning configurations where children can no longer achieve their guessed return values.
Second, the parent functions are \emph{extremal}: they select a single weight rather than computing an aggregate like an average.
The parent objective determines how many guessed values to track.
For $\Inf$ and $\LimInf$, the flattened automaton pairs each child state with a guessed return value, tracking all active children; states with dominated guesses are pruned.
For $\Sup$ and $\LimSup$, exceeding the threshold requires witnessing only one child return value, so we track a guessed value for one distinguished child at a time while other children are tracked only for termination.
This yields substantially smaller state spaces for $\Sup$ and $\LimSup$.

\textit{(4) Limit-average with unbounded children.}
Recall that universality for limit-average QAs is undecidable, and this extends to NQAs with any child aggregator; we therefore focus on emptiness.
For NQAs with $f \in \{\LimInfAvg$, $\LimSupAvg\}$ and $g = \SumMinus$, the construction proceeds in three phases~\cite[Thm.~4.20]{DBLP:journals/tocl/ChatterjeeHO17}.
First, the automaton is determinized via alphabet extension, encoding each combination of nondeterministic choices into a new input letter.
Second, child steps are synchronized so that all active children advance simultaneously.
Third, a powerset-based flattening tracks bounded-multiplicity configurations.
For emptiness of $(\LimSupAvg$, $\SumPlus)$-automata, we first check whether child return values can grow unboundedly; if so, emptiness holds trivially, otherwise the problem reduces to the $\SumBound$ case~\cite[Lem.~5.10]{DBLP:journals/tocl/ChatterjeeHO17}.
The case of $(\LimInfAvg, \SumPlus)$-automata remains open; unlike $\LimSupAvg$, the unboundedness check does not apply since unbounded child returns do not resolve emptiness for infimum-based objectives.

\textit{(5) Silent-weight elimination.}
The procedures above produce QAs with silent weights, arising from silent parent transitions, which invoke no child and therefore emit no returned value.
These must be eliminated to obtain standard QAs.
For extremal objectives, silent weights are replaced by neutral weights (e.g., $-\infty$ for $\Sup$ and $\LimSup$, $+\infty$ for $\Inf$ and $\LimInf$) that do not affect the computed value~\cite[Lem.~4.6]{DBLP:journals/tocl/ChatterjeeHO17}.
For limit-average objectives we can eliminate silent weights via the path-compression construction of~\cite[Lem.~4.7]{DBLP:journals/tocl/ChatterjeeHO17}, which replaces each maximal silent-weight segment by a single shortcut transition, possibly increasing the number of transitions.

\textit{(6) Antichain-based universality.}
We decide universality for NQAs by an on-the-fly search for a counterexample word, pruning the exploration using an antichain as in \textsc{Forklift}~\cite{DBLP:conf/cav/DoveriGM22}.
For the quantitative setting, contexts are enriched with value summaries: in addition to the reachable state set, we store for each state the current maximal weight bound, and extend the subsumption relation to compare both reachability and these bounds~\cite{DBLP:conf/isola/ChalupaHMS24}.
To incorporate B\"uchi acceptance in this version of QuAK, each context keeps two reachability summaries: all reachable pairs, and those reachable via a path that already visited an accepting state.
We only evaluate candidate cycles against the weight threshold once acceptance progress has been witnessed.

\section{QuAK's Architecture and Implementation}
\label{sec:tool}

\begin{figure}[t]
	\centering
	\includegraphics[width=\textwidth,alt={Pipeline of the extended QuAK architecture. The input is parsed into an internal nested quantitative automaton representation, flattened into a quantitative automaton with silent weights, converted by silent-weight elimination, and then passed to existing quantitative automata decision procedures. The parser, internal NQA representation, flattening, and silent-weight elimination components are marked as new.}]{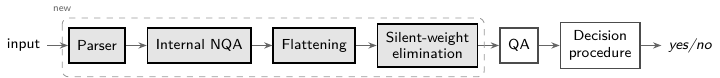}
	\caption{The extended architecture of QuAK. Shaded components are new in this version. A nested quantitative automaton (NQA) is flattened into a standard quantitative automaton (QA) and fed to existing decision procedures.}	
	\label{fig:architecture}
\end{figure}

\subsubsection{Overview}
QuAK~\cite{DBLP:conf/isola/ChalupaHMS24,DBLP:conf/tacas/ChalupaHMS25} is a C++ library and command-line tool for analyzing QAs.
We extend QuAK to support NQAs.
The key addition is a suite of \emph{flattening} procedures that compile an NQA into a QA with silent weights.
This approach supports the known decidable emptiness and universality checks of all combinations of parent objectives in $\{\Sup$, $\Inf$, $\LimSup$, $\LimInf$, $\LimSupAvg$, $\LimInfAvg\}$ with child objectives in $\{\Min$, $\Max$, $\SumBound$, $\SumPlus$, $\SumMinus\}$.
\Cref{fig:architecture} depicts the pipeline; the rest of this section describes each phase and its implementation.

\subsubsection{Input Format and Internal Representation}
QuAK reads automata from plain text files as transition lists \texttt{a : v, q -> p} (letter \texttt{a}, weight \texttt{v}, from state \texttt{q} to \texttt{p}), with the initial state being the source of the first transition.
We extend this format with an optional \texttt{final} directive for Büchi acceptance; otherwise, all states are accepting.
Internally, a QA is stored as an \texttt{Automaton} object that does not fix a value function, allowing the same structure to be analyzed under different objectives.
Automaton states are integer-indexed, and transitions are stored in adjacency-list form with both outgoing and incoming edges.
While constructing an \texttt{Automaton} object, QuAK prunes unreachable states and caches its SCCs~\cite{DBLP:conf/isola/ChalupaHMS24}.
QuAK requires automata to be \emph{complete}: every state must have at least one outgoing transition per letter.
Determinism is not required, but QuAK detects it and dispatches to specialized routines when available.

Input NQAs consist of one \texttt{@PARENT} block and one \texttt{@CHILD}~$i$ block per child.
Children use the syntax above, while parent transitions use \texttt{a : i, q -> p}, where $i > 0$ invokes child~$i$ and $i = 0$ denotes a silent transition.
The parent uses Büchi acceptance, and children use finite-word acceptance.
Internally, the \texttt{NestedAutomaton} and \texttt{ChildAutomaton} classes extend QuAK's base \texttt{Automaton} class.
A \texttt{NestedAutomaton} object has a parent automaton of type \texttt{Automaton} and a vector of pointers to \texttt{ChildAutomaton} objects.

\subsubsection{Flattening Nested Quantitative Automata}
The flattening module implements the constructions; see~\Cref{sec:theory} for the algorithmic details.

\textit{(1) Regular children.}
For NQAs with $g \in \{\Min,\Max,\SumBound\}$, QuAK implements an obligation-based flattening that preserves semantic equivalence.
When a parent transition invokes a child, the flattened automaton guesses the child's return value on that transition; the guess is verified by an obligation tracking possible child configurations in parallel.
Each active obligation records the child index, the guessed value, and a frontier of $(\textit{childState}, \textit{accumulatedValue})$ pairs; configurations that cannot reach a final state with the guessed value are pruned via target-aware reachability tables from reverse BFS.
Return-value sets are computed parent-aware by exploring the synchronized parent--child product restricted to parent states that can still reach an accepting SCC.
QuAK applies this procedure to universality for $f \in \{\Inf$, $\Sup$, $\LimInf$, $\LimSup\}$ and $g \in \{\Min$, $\Max$, $\SumBound\}$, and to emptiness when $g=\SumBound$ or when $f\in\{\LimInfAvg$, $\LimSupAvg\}$ and $g\in\{\Min$, $\Max\}$; remaining emptiness cases use specialized procedures.

\textit{(2) Threshold-based bounding.}
For universality of NQAs with $f \in \{\Sup$, $\LimSup$, $\Inf$, $\LimInf\}$ and $g \in \{\SumPlus$, $\SumMinus\}$, QuAK reduces the unbounded child aggregator to $\SumBound$ via threshold-preserving clipping:
values at or above the threshold collapse to it for $\SumPlus$; values below collapse to a strictly smaller value for $\SumMinus$.
The resulting bounded-return automaton is handled by the procedure above.
The same reduction applies semantically to the corresponding emptiness cases, but QuAK uses the specialized threshold constructions below.

\textit{(3) Extremal parents with monotonic children.}
For emptiness with $f\in\{\Sup$, $\LimSup$, $\Inf$, $\LimInf\}$ and $g\in\{\Min$, $\Max$, $\SumPlus$, $\SumMinus\}$, QuAK avoids computing exact child return values.
Instead, each child invocation is reduced to a binary threshold outcome: whether the child can return a value at least $\lambda$.
For $\Min$ and $\Max$, the verifier only needs to remember the current below/above-threshold status of each tracked child run; for $\SumPlus$ and $\SumMinus$, it additionally stores capped threshold progress.
The implementation precomputes backward liveness information from final child states and uses it to prune guesses that can no longer realize their threshold outcome.
The flattened automaton therefore has weights in $\{0,1\}$, and emptiness is checked with the same parent aggregator $f$ against threshold $1$.
For $f\in\{\Sup,\LimSup\}$, QuAK uses a single-witness variant: the flattened state stores one explicitly tracked threshold-reaching child, while the remaining active children are represented only by termination obligations.

\textit{(4) Limit-average with unbounded children.}
For $(\LimSupAvg,\SumPlus)$ emptiness, QuAK separates the unbounded case from the bounded one.
It computes a threshold $\lambda$ capturing the largest value a single child can accumulate without forcing a repetition of some global configuration with a strictly larger accumulated sum, and checks whether the instance is nonempty against $\lambda$.
If this check succeeds, the instance is nonempty; otherwise the relevant child values are bounded, and QuAK calls the regular $\SumBound$ flattening with $B=\lambda$.
For $(\LimInfAvg,\SumMinus)$ and $(\LimSupAvg,\SumMinus)$ emptiness, QuAK first prepares the input for the synchronization construction, completing or determinizing via alphabet extension, it when needed.
It then replaces all children by one synchronized ultimate child and flattens by tracking a bounded multiset of active ultimate-child states.
The multiset is stored sparsely as sorted $(\textit{stateId},\textit{count})$ pairs; BFS generates states on demand and drops successors whose multiplicity exceeds the computed bound $c$.

\textit{(5) Silent-weight elimination.}
Before passing the flattened automaton to the existing decision procedures, QuAK eliminates silent-weight transitions.
For prefix-independent objectives ($\LimSup$, $\LimInf$, $\LimSupAvg$, $\LimInfAvg$), QuAK reduces the potential overhead by only considering those within accepting SCCs.
This optimization is enabled by default, but can be deactivated.

\textit{(6) Antichains, interfacing, and optimizations.}
The flattening routines are used internally by the nested emptiness and universality checks, and exposed through the public API for custom analysis pipelines.
For limit-average emptiness of QAs with B\"uchi acceptance, QuAK computes maximum mean cycles separately within accepting SCCs, ignoring non-accepting ones.
For \textsc{Forklift} inclusion~\cite{DBLP:conf/cav/DoveriGM22}, contexts now store two relations per weight level: all reachable target pairs, and the subset reachable via a path through an accepting state.
The lasso membership test was likewise adjusted to require an accepting visit on the target cycle, ensuring inclusion counterexamples respect B\"uchi acceptance rather than only the weight threshold.

\subsubsection{Availability and Usage}
QuAK is open-source under the MIT license and available online.\footnote[1]{\url{https://github.com/ista-vamos/nested-quak}}
The tool has no external dependencies other than a C++17 compiler.
Instructions and examples are provided in the repository's README.

\section{Experimental Evaluation}
\label{sec:experiments}

We evaluate QuAK's new capabilities along two benchmarks: (i) \emph{response-time} properties and (ii) \emph{resource-consumption} constraints.
The experiments are designed to isolate the main sources of blowup in the flattening constructions: child return-value range, number of simultaneously active children, child-state space, and transition density of the flattened automaton.

The response-time benchmark is based on the standard example from~\cite{DBLP:journals/tocl/ChatterjeeHO17}, with parameters added to control the number of simultaneously pending requests and the response-time bound.
The resource-consumption benchmark follows the construction of~\cite[Thm.~6.2]{DBLP:journals/tocl/ChatterjeeHO17}, where it is used to show an exponential succinctness gap between deterministic NQAs and nondeterministic QAs.
The generator scripts are included with the artifact.

\subsubsection{Setup}
All experiments were run on Ubuntu 24.04.3 with an Intel Ultra 7 Processor 255U and 32~GB RAM.
The tool was compiled with GCC using C++17 and optimization level \texttt{-O3}.
Each instance was limited to 300~seconds and 30~GB of memory.
Reported runtimes are wall-clock seconds averaged over three runs, excluding parsing time; each measurement was preceded by a warm-up run.

\subsection{Response Time}
\label{sec:response}

\begin{figure}[t]
	\centering
	\includegraphics[scale=1,alt={Example response-time benchmark automaton A two two. The parent automaton tracks at most two pending requests, rejects when the bound is violated, and spawns a two-state child automaton on requests. The child counts elapsed positions until a grant is read.}]{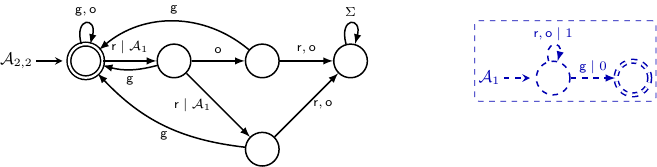}
	\caption{The response-time automaton $\AA_{2,2} = (\A_{2,2}, \A_1)$. Every parent transition without a child automaton annotation is silent.}
	\label{fig:example_response}
\end{figure}

We study a parametric family of NQAs $\AA_{n,k}$ for bounded-response properties in request-grant protocols.
The parameter $n$ bounds the number of simultaneously pending requests, and $k$ bounds the allowable response time.
The alphabet comprises a request~$\mathsf{r}$, a grant~$\mathsf{g}$, and a neutral action~$\mathsf{o}$.
For $n \geq 1$ and $k \geq n$, the parent automaton has $\frac{n(2k-n+1)}{2} + 2$ states and uses a single two-state child automaton.
A child instance is spawned on each request, so at most $n$ instances are concurrently active; if this bound is exceeded or some request goes ungranted for more than $k$ steps, the parent moves to a rejecting sink.
Each child accumulates the number of steps between its spawning request and the matching grant, and the parent value function determines how these per-request response times are combined: $\Sup$ yields the worst-case response time, while $\LimSupAvg$ yields the long-run average.
\Cref{fig:example_response} shows a representative instance $\AA_{2,2}$.

Unlike the unbounded response-time automaton in~\Cref{fig:response_time_intro}, the bounds enforced by $\AA_{n,k}$ yield a finite-state benchmark suitable for systematic experiments.
This family also captures a standard verification scenario: given a finite-state model of a server, quantify the worst-case or average response time over all permitted behaviors.
We evaluate QuAK on this benchmark with emptiness of $(\Sup,\SumPlus)$-automata, emptiness of $(\LimSupAvg,\SumPlus)$-automata, and universality of $(\Sup,\SumBound)$-automata.
The results are reported in \Cref{tab:response_time_benchmarks}.

\begin{table}[t]
	\centering
	\caption{Response time benchmarks: runtime in seconds, \oom~= out of memory (30\,GB), \oot~= out of time (300\,s).} \label{tab:response_time_benchmarks}
	\begin{adjustbox}{max width=\linewidth}
		\scriptsize
		\setlength{\tabcolsep}{2pt}
		\renewcommand{\arraystretch}{1.05}
		\begin{tabular}{@{}c@{\qquad}c@{\qquad}c@{}}
			\begin{tabular}[t]{@{}r*{8}{c}@{}}
				\multicolumn{9}{c}{\makebox[0pt]{(a) (\Sup,\SumPlus) emptiness}}\\[2pt]
				\toprule
				$n\backslash k$ & 4 & 8 & 16 & 32 & 64 & 128 & 256 & 512 \\
				\midrule
				4 & .00 & .00 & .01 & .01 & .06 & .32 & 1.5 & 6.3 \\
				8 & -- & .00 & .01 & .03 & .16 & .76 & 3.3 & 14 \\
				16 & -- & -- & .01 & .07 & .39 & 1.7 & 7.3 & 32 \\
				32 & -- & -- & -- & .13 & .85 & 3.8 & 17 & 72 \\
				64 & -- & -- & -- & -- & 1.3 & 7.8 & 35 & \oom \\
				128 & -- & -- & -- & -- & -- & 12 & 71 & \oom \\
				256 & -- & -- & -- & -- & -- & -- & \oom & \oom \\
				512 & -- & -- & -- & -- & -- & -- & -- & \oom \\
				\bottomrule
			\end{tabular}
			&
			\begin{tabular}[t]{@{}r*{8}{c}@{}}
				\multicolumn{9}{c}{\makebox[0pt]{(b) (\LimSupAvg,\SumPlus) emptiness}}\\[2pt]
				\toprule
				$n\backslash k$ & 3 & 4 & 5 & 6 & 7 & 8 & 9 & 10 \\
				\midrule
				3 & .00 & .00 & .00 & .01 & .02 & .05 & .13 & .31 \\
				4 & -- & .00 & .00 & .01 & .07 & .28 & 1.1 & 3.2 \\
				5 & -- & -- & .00 & .03 & .21 & 1.3 & 5.5 & 20 \\
				6 & -- & -- & -- & .03 & .29 & 2.8 & 18 & 93 \\
				7 & -- & -- & -- & -- & .33 & 4.3 & 39 & 284 \\
				8 & -- & -- & -- & -- & -- & 4.7 & 58 & \oom \\
				9 & -- & -- & -- & -- & -- & -- & 63 & \oom \\
				10 & -- & -- & -- & -- & -- & -- & -- & \oom \\
				\bottomrule
			\end{tabular}
			&
			\begin{tabular}[t]{@{}r*{8}{c}@{}}
				\multicolumn{9}{c}{\makebox[0pt]{(c) (\Sup,\SumBound) universality}}\\[2pt]
				\toprule
				$n\backslash k$ & 2 & 3 & 4 & 5 & 6 & 7 & 8 & 9 \\
				\midrule
				2 & .00 & .00 & .01 & .04 & .14 & .42 & 1.2 & 3.6 \\
				3 & -- & .00 & .02 & .08 & .40 & 2.3 & 12 & 44 \\
				4 & -- & -- & .02 & .12 & 1.1 & 12 & 159 & \oot \\
				5 & -- & -- & -- & .14 & 2.0 & 51 & \oot & \oot \\
				6 & -- & -- & -- & -- & 2.4 & 136 & \oot & \oom \\
				7 & -- & -- & -- & -- & -- & 177 & \oot & \oom \\
				8 & -- & -- & -- & -- & -- & -- & \oot & \oom \\
				9 & -- & -- & -- & -- & -- & -- & -- & \oom \\
				\bottomrule
			\end{tabular}
		\end{tabular}
	\end{adjustbox}
\end{table}

\subsubsection{Emptiness of $(\Sup,\SumPlus)$-automata}
\Cref{tab:response_time_benchmarks}a shows that emptiness checking scales well on this family: the dependence on $k$ is close to quadratic, while the dependence on $n$ is comparatively mild.
This behavior is consistent with the specialized procedure for extremal parents with monotonic children (\Cref{sec:theory}) and with the structure of $\AA_{n,k}$.
All children are copies of the same deterministic two-state automaton: before the next grant, every active child is in the same nonfinal state and differs only by accumulated $\SumPlus$ value.
The specialized construction stores capped threshold progress only for one distinguished witness; the other active children contribute identical termination obligations, so increasing $n$ mainly enlarges the parent queue state space.

\subsubsection{Emptiness of $(\LimSupAvg,\SumPlus)$-automata}
\Cref{tab:response_time_benchmarks}b shows a different pattern: runtime grows quickly with both parameters, with the feasible range of $n$ much smaller than in \Cref{tab:response_time_benchmarks}a.
For these bounded-response instances, the unbounded-child case is ruled out and QuAK falls back to regular $\SumBound$ flattening.
Each spawned child is represented by a guessed response time and an obligation verifying the guess until the next grant.
There are $O(k)$ possible guesses per child, and up to $n$ such obligations may be live simultaneously.
Thus $k$ enlarges the per-child value domain, while $n$ determines the combinatorial search space by increasing the number of active guesses that must be combined.

\subsubsection{Universality of $(\Sup,\SumBound)$-automata}
\Cref{tab:response_time_benchmarks}c shows that universality follows a similar pattern: runtime grows rapidly in both parameters and reaches time or memory limits at smaller values than in the emptiness experiments.
The source of growth is the same: regular $\SumBound$ flattening combines $O(k)$ response-time guesses across up to $n$ concurrently active children.
Universality is consistently more expensive because handling the flattened automaton's nondeterminism requires a \PSpace procedure rather than the \PTime emptiness check.

\subsection{Resource Consumption}
\label{sec:resource}

\begin{figure}[t]
	\centering
	\includegraphics[scale=1,alt={Example resource-consumption benchmark automaton B one two. The parent spawns a child when the single process starts. The child tracks which of two resources have been accessed and returns zero, one, or two when the process terminates. Silent parent transitions and zero-weight child transitions are drawn without weight annotations; the child rejecting sink is omitted.}]{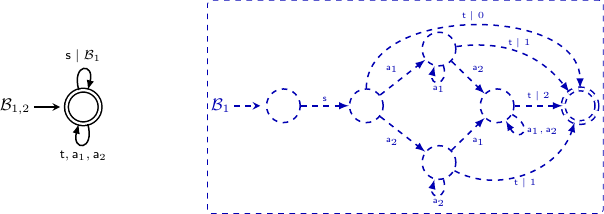}
	\caption{The resource-consumption automaton $\BB_{1,2} = (\B_{1,2}, \B_1)$. Every parent transition without a child automaton annotation is silent. The rejecting sink state of the child automaton and transitions to it are not shown. Every child transition without a weight annotation has weight 0.}
	\label{fig:example_resource}
\end{figure}

We study another parametric family of NQAs $\BB_{n,k}$ that monitors resource consumption in a system with dynamically started and terminated processes.
The parameter $n$ bounds the number of concurrently running processes, and $k$ bounds the number of distinct resources.
The alphabet comprises $n(k+2)$ actions: for each process $i \in \{1,\ldots,n\}$, a start action $s_i$, a termination action $t_i$, and $k$ resource-access actions $a_{i,1}, \ldots, a_{i,k}$.
For $n \geq 1$ and $k \geq 1$, the parent automaton has a single control state and uses $n$ child templates $B_1, \ldots, B_n$, each with $2^k + 3$ states.
When process $i$ starts, the parent spawns an instance of child $B_i$, which remains active until the matching termination $t_i$.
While active, the instance tracks which resources process $i$ accesses; upon termination, it returns the number of distinct resources used, computed via the $\Max$ value function.
The child rejects the words that violate the protocol, such as starting an already-running process.
The parent aggregates per-process values with either $\Sup$, yielding the maximal consumption across executions, or $\LimSupAvg$, yielding the long-run average.
\Cref{fig:example_resource} shows a representative instance $\BB_{1,2}$.

This family provides a controlled benchmark with two orthogonal scaling axes: increasing $k$ multiplies the resource-usage patterns a process can exhibit, while increasing $n$ multiplies concurrent process overlap.
Although $\BB_{n,k}$ is deterministic as a nested automaton, any equivalent non-nested automaton must encode joint resource subsets explicitly and can therefore grow exponentially in the number of concurrently tracked processes~\cite{DBLP:journals/tocl/ChatterjeeHO17}.
We evaluate QuAK on emptiness of $(\Sup,\Max)$- and $(\LimSupAvg,\Max)$-automata; results appear in \Cref{tab:resource_consumption_benchmarks}.

\begin{table}[t]
	\centering
	\caption{Resource consumption benchmarks: runtime in seconds, \oom~= out of memory (30\,GB), \oot~= out of time (300\,s).} \label{tab:resource_consumption_benchmarks}
	\begin{adjustbox}{max width=\linewidth}
		\scriptsize
		\setlength{\tabcolsep}{2pt}
		\renewcommand{\arraystretch}{1.05}
		\begin{tabular}{@{}c@{\qquad}c@{}}
			\begin{tabular}[t]{@{}r*{6}{c}@{}}
				\multicolumn{7}{c}{\makebox[0pt]{(a) (\Sup,\Max) emptiness}}\\[2pt]
				\toprule
				$n\backslash k$ & 1 & 2 & 3 & 4 & 5 & 6 \\
				\midrule
				1 & .00 & .00 & .00 & .00 & .00 & .00 \\
				2 & .00 & .00 & .01 & .06 & .37 & 1.9 \\
				3 & .01 & .10 & 1.1 & 11 & 103 & \oom \\
				4 & .12 & 2.4 & 43 & \oom & \oom & \oom \\
				5 & 1.3 & 38 & \oom & \oom & \oom & \oom \\
				6 & 9.9 & \oom & \oom & \oom & \oom & \oom \\
				\bottomrule
			\end{tabular}
			&\hspace{4em}
			\begin{tabular}[t]{@{}r*{4}{c}@{}}
				\multicolumn{5}{c}{\makebox[0pt]{(b) (\LimSupAvg,\Max) emptiness}}\\[2pt]
				\toprule
				$n\backslash k$ & 1 & 2 & 3 & 4 \\
				\midrule
				1 & .00 & .00 & .00 & .00 \\
				2 & .00 & .03 & 1.1 & 62 \\
				3 & .12 & 48 & \oot & \oot \\
				4 & 20 & \oot & \oot & \oom \\
				\bottomrule
			\end{tabular}
		\end{tabular}
	\end{adjustbox}
\end{table}

\subsubsection{Emptiness of $(\Sup,\Max)$-automata}
\Cref{tab:resource_consumption_benchmarks}a shows rapid blowup in both parameters.
Each child encodes which resource subset a process has accessed, giving $\Theta(2^k)$ control states per process.
The $\Sup$ construction tracks one witness (whether some process sees all resources) but still carries termination obligations for the remaining children.
Since a run may have one active instance per process template, these obligations yield a $2^{\Theta(nk)}$ joint space.

\subsubsection{Emptiness of $(\LimSupAvg,\Max)$-automata}
As \Cref{tab:resource_consumption_benchmarks}b shows, the limit-average case becomes substantially harder than the supremum once multiple processes overlap.
The $\Sup$ procedure needs only one threshold-reaching witness, whereas the $\LimSupAvg$ regular-children procedure must preserve the exact return values of all active children.
These returns range over $0,\ldots,k$, and each verifier still carries its child's resource-subset state.
Combined with the $2^{\Theta(nk)}$ subset space, this exceeds resource budgets at small parameter values.

\section{Conclusion}
\label{sec:conclusion}

We presented the first tool for analyzing nested quantitative automata (NQAs), extending QuAK with flattening procedures that reduce nested emptiness and universality to their non-nested counterparts.
This enables automated verification of quantitative properties beyond standard quantitative automata, such as average response time.
Our experiments show that explicit flattening can incur large state spaces, especially when many child instances are active.
Symbolic state representations~\cite{DBLP:journals/iandc/BurchCMDH92} and domain-specific reductions~\cite{DBLP:conf/kbse/BokorKSS11} could mitigate this blowup.
Identifying tractable fragments, informed for example by safety-liveness classifications~\cite{DBLP:conf/fossacs/HenzingerMS23,DBLP:conf/concur/BokerHMS23,DBLP:journals/lmcs/BokerHMS25}, would broaden the tool's applicability.

\begin{credits}
\subsubsection{\ackname}
This work was supported by the European Research Council (ERC) Grants VAMOS (No. 101020093) and HYPER (No. 101055412).
\subsubsection{\discintname}
The authors have no competing interests to declare.
\end{credits}

\section*{Data-Availability Statement}

The artifact supporting the experimental results in this paper is available in the QuAK repository at \url{https://github.com/ista-vamos/nested-quak}.
It contains the extended QuAK implementation, benchmark generators, example inputs, and scripts/logs for reproducing the reported tables.
The artifact is intended to reproduce the experiments under the setup described in Section~\ref{sec:experiments}; runtimes may vary across machines, and the reported timeout and memory-exhaustion results depend on the stated hardware limits.
No sensitive or restricted data are used.
An archived version is available on Zenodo at DOI: \url{http://doi.org/10.5281/zenodo.19844606}.

\bibliographystyle{splncs04}
\bibliography{main}

@article{DBLP:journals/iandc/BurchCMDH92,
  author    = {Jerry R. Burch and
               Edmund M. Clarke and
               Kenneth L. McMillan and
               David L. Dill and
               L. J. Hwang},
  title     = {Symbolic Model Checking: 10{\^{}}20 States and Beyond},
  journal   = {Information and Computation},
  volume    = {98},
  number    = {2},
  pages     = {142--170},
  year      = {1992},
  doi       = {10.1016/0890-5401(92)90017-A}
}

@inproceedings{DBLP:conf/kbse/BokorKSS11,
  author    = {P{\'e}ter Bokor and
               Johannes Kinder and
               Marco Serafini and
               Neeraj Suri},
  title     = {Supporting Domain-Specific State Space Reductions through
               Local Partial-Order Reduction},
  booktitle = {26th {IEEE/ACM} International Conference on Automated Software
               Engineering, {ASE} 2011},
  pages     = {113--122},
  publisher = {{IEEE}},
  year      = {2011},
  doi       = {10.1109/ASE.2011.6100044}
}

@article{DBLP:journals/lmcs/BokerHMS25,
  author       = {Udi Boker and
                  Thomas A. Henzinger and
                  Nicolas Mazzocchi and
                  N. Ege Sara{\c{c}}},
  title        = {Safety and Liveness of Quantitative Properties and Automata},
  journal      = {Log. Methods Comput. Sci.},
  volume       = {21},
  number       = {2},
  year         = {2025},
  url          = {https://doi.org/10.46298/lmcs-21(2:2)2025},
  doi          = {10.46298/LMCS-21(2:2)2025},
  bibsource    = {dblp computer science bibliography, https://dblp.org}
}

@article{DBLP:journals/sttt/HenselJKQV22,
  author       = {Christian Hensel and
                  Sebastian Junges and
                  Joost{-}Pieter Katoen and
                  Tim Quatmann and
                  Matthias Volk},
  title        = {The probabilistic model checker Storm},
  journal      = {Int. J. Softw. Tools Technol. Transf.},
  volume       = {24},
  number       = {4},
  pages        = {589--610},
  year         = {2022},
  url          = {https://doi.org/10.1007/s10009-021-00633-z},
  doi          = {10.1007/S10009-021-00633-Z},
  bibsource    = {dblp computer science bibliography, https://dblp.org}
}

@inproceedings{DBLP:conf/mfcs/ChatterjeeHO16,
  author       = {Krishnendu Chatterjee and
                  Thomas A. Henzinger and
                  Jan Otop},
  editor       = {Piotr Faliszewski and
                  Anca Muscholl and
                  Rolf Niedermeier},
  title        = {Nested Weighted Limit-Average Automata of Bounded Width},
  booktitle    = {41st International Symposium on Mathematical Foundations of Computer
                  Science, {MFCS} 2016, Krak{\'{o}}w, Poland, August 22-26, 2016},
  series       = {LIPIcs},
  volume       = {58},
  pages        = {24:1--24:14},
  publisher    = {Schloss Dagstuhl - Leibniz-Zentrum f{\"{u}}r Informatik},
  year         = {2016},
  url          = {https://doi.org/10.4230/LIPIcs.MFCS.2016.24},
  doi          = {10.4230/LIPICS.MFCS.2016.24},
  bibsource    = {dblp computer science bibliography, https://dblp.org}
}

@inproceedings{DBLP:conf/sas/ChatterjeeHO16,
  author       = {Krishnendu Chatterjee and
                  Thomas A. Henzinger and
                  Jan Otop},
  editor       = {Xavier Rival},
  title        = {Quantitative Monitor Automata},
  booktitle    = {Static Analysis - 23rd International Symposium, {SAS} 2016, Edinburgh,
                  UK, September 8-10, 2016, Proceedings},
  series       = {Lecture Notes in Computer Science},
  volume       = {9837},
  pages        = {23--38},
  publisher    = {Springer},
  year         = {2016},
  url          = {https://doi.org/10.1007/978-3-662-53413-7\_2},
  doi          = {10.1007/978-3-662-53413-7\_2},
  bibsource    = {dblp computer science bibliography, https://dblp.org}
}

@inproceedings{DBLP:conf/tacas/ChalupaHMS25,
  author       = {Marek Chalupa and
                  Thomas A. Henzinger and
                  Nicolas Mazzocchi and
                  N. Ege Sara{\c{c}}},
  editor       = {Arie Gurfinkel and
                  Marijn Heule},
  title        = {Automating the Analysis of Quantitative Automata with QuAK},
  booktitle    = {Tools and Algorithms for the Construction and Analysis of Systems
                  - 31st International Conference, {TACAS} 2025, Held as Part of the
                  International Joint Conferences on Theory and Practice of Software,
                  {ETAPS} 2025, Hamilton, ON, Canada, May 3-8, 2025, Proceedings, Part
                  {I}},
  series       = {Lecture Notes in Computer Science},
  volume       = {15696},
  pages        = {303--312},
  publisher    = {Springer},
  year         = {2025},
  url          = {https://doi.org/10.1007/978-3-031-90643-5\_16},
  doi          = {10.1007/978-3-031-90643-5\_16},
  bibsource    = {dblp computer science bibliography, https://dblp.org}
}

@inproceedings{DBLP:conf/isola/ChalupaHMS24,
  author       = {Marek Chalupa and
                  Thomas A. Henzinger and
                  Nicolas Mazzocchi and
                  N. Ege Sara{\c{c}}},
  editor       = {Tiziana Margaria and
                  Bernhard Steffen},
  title        = {QuAK: Quantitative Automata Kit},
  booktitle    = {Leveraging Applications of Formal Methods, Verification and Validation.
                  Software Engineering Methodologies - 12th International Symposium,
                  ISoLA 2024, Crete, Greece, October 27-31, 2024, Proceedings, Part
                  {IV}},
  series       = {Lecture Notes in Computer Science},
  volume       = {15222},
  pages        = {3--20},
  publisher    = {Springer},
  year         = {2024},
  url          = {https://doi.org/10.1007/978-3-031-75387-9\_1},
  doi          = {10.1007/978-3-031-75387-9\_1},
  bibsource    = {dblp computer science bibliography, https://dblp.org}
}

@inproceedings{DBLP:conf/wia/DemailleDLS13,
  author       = {Akim Demaille and
                  Alexandre Duret{-}Lutz and
                  Sylvain Lombardy and
                  Jacques Sakarovitch},
  editor       = {Stavros Konstantinidis},
  title        = {Implementation Concepts in Vaucanson 2},
  booktitle    = {Implementation and Application of Automata - 18th International Conference,
                  {CIAA} 2013, Halifax, NS, Canada, July 16-19, 2013. Proceedings},
  series       = {Lecture Notes in Computer Science},
  volume       = {7982},
  pages        = {122--133},
  publisher    = {Springer},
  year         = {2013},
  doi          = {10.1007/978-3-642-39274-0\_12},
  bibsource    = {dblp computer science bibliography, https://dblp.org}
}

@manual{Awali2.3,
  title = {Awali, a library for weighted automata and transducers (version 2.3)},
  author = {Sylvain Lombardy and Victor Marsault and Jacques Sakarovitch},
  year = {2022},
  note = {Software available at {http://vaucanson-project.org/Awali/2.3/}},
}

@inproceedings{DBLP:conf/wia/LombardyPRS03,
  author       = {Sylvain Lombardy and
                  Raphael Poss and
                  Yann R{\'{e}}gis{-}Gianas and
                  Jacques Sakarovitch},
  editor       = {Oscar H. Ibarra and
                  Zhe Dang},
  title        = {Introducing {VAUCANSON}},
  booktitle    = {Implementation and Application of Automata, 8th International Conference,
                  {CIAA} 2003, Santa Barbara, California, USA, July 16-18, 2003, Proceedings},
  series       = {Lecture Notes in Computer Science},
  volume       = {2759},
  pages        = {96--107},
  publisher    = {Springer},
  year         = {2003},
  doi          = {10.1007/3-540-45089-0\_10},
  bibsource    = {dblp computer science bibliography, https://dblp.org}
}

@inproceedings{DBLP:conf/cpe/KwiatkowskaNP02,
  author       = {Marta Z. Kwiatkowska and
                  Gethin Norman and
                  David Parker},
  editor       = {Tony Field and
                  Peter G. Harrison and
                  Jeremy T. Bradley and
                  Uli Harder},
  title        = {{PRISM:} Probabilistic Symbolic Model Checker},
  booktitle    = {Computer Performance Evaluation, Modelling Techniques and Tools 12th
                  International Conference, {TOOLS} 2002, London, UK, April 14-17, 2002,
                  Proceedings},
  series       = {Lecture Notes in Computer Science},
  volume       = {2324},
  pages        = {200--204},
  publisher    = {Springer},
  year         = {2002},
  doi          = {10.1007/3-540-46029-2\_13},
  bibsource    = {dblp computer science bibliography, https://dblp.org}
}

@inproceedings{DBLP:conf/atva/Nickovic020,
  author       = {Dejan Nickovic and
                  Tomoya Yamaguchi},
  editor       = {Dang Van Hung and
                  Oleg Sokolsky},
  title        = {{RTAMT:} Online Robustness Monitors from {STL}},
  booktitle    = {Automated Technology for Verification and Analysis - 18th International
                  Symposium, {ATVA} 2020, Hanoi, Vietnam, October 19-23, 2020, Proceedings},
  series       = {Lecture Notes in Computer Science},
  volume       = {12302},
  pages        = {564--571},
  publisher    = {Springer},
  year         = {2020},
  doi          = {10.1007/978-3-030-59152-6\_34},
  bibsource    = {dblp computer science bibliography, https://dblp.org}
}

@inproceedings{DBLP:conf/cav/Donze10,
  author       = {Alexandre Donz{\'{e}}},
  editor       = {Tayssir Touili and
                  Byron Cook and
                  Paul B. Jackson},
  title        = {Breach, {A} Toolbox for Verification and Parameter Synthesis of Hybrid
                  Systems},
  booktitle    = {Computer Aided Verification, 22nd International Conference, {CAV}
                  2010, Edinburgh, UK, July 15-19, 2010. Proceedings},
  series       = {Lecture Notes in Computer Science},
  volume       = {6174},
  pages        = {167--170},
  publisher    = {Springer},
  year         = {2010},
  doi          = {10.1007/978-3-642-14295-6\_17},
  bibsource    = {dblp computer science bibliography, https://dblp.org}
}

@inproceedings{DBLP:conf/tacas/AnnpureddyLFS11,
  author       = {Yashwanth Annpureddy and
                  Che Liu and
                  Georgios Fainekos and
                  Sriram Sankaranarayanan},
  editor       = {Parosh Aziz Abdulla and
                  K. Rustan M. Leino},
  title        = {S-TaLiRo: {A} Tool for Temporal Logic Falsification for Hybrid Systems},
  booktitle    = {Tools and Algorithms for the Construction and Analysis of Systems
                  - 17th International Conference, {TACAS} 2011, Held as Part of the
                  Joint European Conferences on Theory and Practice of Software, {ETAPS}
                  2011, Saarbr{\"{u}}cken, Germany, March 26-April 3, 2011. Proceedings},
  series       = {Lecture Notes in Computer Science},
  volume       = {6605},
  pages        = {254--257},
  publisher    = {Springer},
  year         = {2011},
  doi          = {10.1007/978-3-642-19835-9\_21},
  bibsource    = {dblp computer science bibliography, https://dblp.org}
}

@inproceedings{DBLP:conf/hybrid/HenzingerH94a,
  author       = {Thomas A. Henzinger and
                  Pei{-}Hsin Ho},
  editor       = {Panos J. Antsaklis and
                  Wolf Kohn and
                  Anil Nerode and
                  Shankar Sastry},
  title        = {{HYTECH:} The Cornell HYbrid TECHnology Tool},
  booktitle    = {Hybrid Systems II, Proceedings of the Third International Workshop
                  on Hybrid Systems, Ithaca, NY, USA, October 1994},
  series       = {Lecture Notes in Computer Science},
  volume       = {999},
  pages        = {265--293},
  publisher    = {Springer},
  year         = {1994},
  doi          = {10.1007/3-540-60472-3\_14},
  bibsource    = {dblp computer science bibliography, https://dblp.org}
}

@article{DBLP:journals/sttt/LarsenPY97,
  author       = {Kim Guldstrand Larsen and
                  Paul Pettersson and
                  Wang Yi},
  title        = {{UPPAAL} in a Nutshell},
  journal      = {Int. J. Softw. Tools Technol. Transf.},
  volume       = {1},
  number       = {1-2},
  pages        = {134--152},
  year         = {1997},
  doi          = {10.1007/S100090050010},
  bibsource    = {dblp computer science bibliography, https://dblp.org}
}

@inproceedings{DBLP:conf/cav/DoveriGM22,
  author    = {Kyveli Doveri and
               Pierre Ganty and
               Nicolas Mazzocchi},
  editor    = {Sharon Shoham and
               Yakir Vizel},
  title     = {FORQ-Based Language Inclusion Formal Testing},
  booktitle = {Computer Aided Verification - 34th International Conference, {CAV}
               2022, Haifa, Israel, August 7-10, 2022, Proceedings, Part {II}},
  series    = {Lecture Notes in Computer Science},
  volume    = {13372},
  pages     = {109--129},
  publisher = {Springer},
  year      = {2022},
  doi       = {10.1007/978-3-031-13188-2\_6},
  bibsource = {dblp computer science bibliography, https://dblp.org}
}

@misc{forklift,
	author = {
	Kyveli Doveri and 
	Pierre Ganty and
	Nicolas Mazzocchi},
	year = {2022}, 
	title = {{FORKLIFT} (v1.0)},
	howpublished = {Zenodo},
	doi = {10.5281/zenodo.6552870},
	note = {maintained at https://github.com/Mazzocchi/FORKLIFT},
}

@article{DBLP:journals/tocl/ChatterjeeDH10,
	author       = {Krishnendu Chatterjee and
	                Laurent Doyen and
	                Thomas A. Henzinger},
	title        = {Quantitative languages},
	journal      = {{ACM} Trans. Comput. Log.},
	volume       = {11},
	number       = {4},
	pages        = {23:1--23:38},
	year         = {2010},
	doi          = {10.1145/1805950.1805953},
	bibsource    = {dblp computer science bibliography, https://dblp.org}
}

@article{DBLP:journals/tocl/ChatterjeeHO17,
  author    = {Krishnendu Chatterjee and
               Thomas A. Henzinger and
               Jan Otop},
  title     = {Nested Weighted Automata},
  journal   = {{ACM} Trans. Comput. Log.},
  volume    = {18},
  number    = {4},
  pages     = {31:1--31:44},
  year      = {2017},
  doi       = {10.1145/3152769},
  bibsource = {dblp computer science bibliography, https://dblp.org}
}

@inproceedings{DBLP:conf/fossacs/HenzingerMS23,
  author       = {Thomas A. Henzinger and
                  Nicolas Mazzocchi and
                  N. Ege Sara{\c{c}}},
  editor       = {Orna Kupferman and
                  Pawel Sobocinski},
  title        = {Quantitative Safety and Liveness},
  booktitle    = {Foundations of Software Science and Computation Structures - 26th
                  International Conference, FoSSaCS 2023, Held as Part of the European
                  Joint Conferences on Theory and Practice of Software, {ETAPS} 2023,
                  Paris, France, April 22-27, 2023, Proceedings},
  series       = {Lecture Notes in Computer Science},
  volume       = {13992},
  pages        = {349--370},
  publisher    = {Springer},
  year         = {2023},
  doi          = {10.1007/978-3-031-30829-1\_17},
  bibsource    = {dblp computer science bibliography, https://dblp.org}
}

@inproceedings{DBLP:conf/concur/BokerHMS23,
  author       = {Udi Boker and
                  Thomas A. Henzinger and
                  Nicolas Mazzocchi and
                  N. Ege Sara{\c{c}}},
  editor       = {Guillermo A. P{\'{e}}rez and
                  Jean{-}Fran{\c{c}}ois Raskin},
  title        = {Safety and Liveness of Quantitative Automata},
  booktitle    = {34th International Conference on Concurrency Theory, {CONCUR} 2023,
                  September 18-23, 2023, Antwerp, Belgium},
  series       = {LIPIcs},
  volume       = {279},
  pages        = {17:1--17:18},
  publisher    = {Schloss Dagstuhl - Leibniz-Zentrum f{\"{u}}r Informatik},
  year         = {2023},
  doi          = {10.4230/LIPICS.CONCUR.2023.17},
  bibsource    = {dblp computer science bibliography, https://dblp.org}
}

\end{document}